\documentclass[acus]{JAC2000}
\usepackage{times}
\usepackage{graphicx}

\begin{document}
\title{OPTIMIZED WAKEFIELD COMPUTATIONS USING A NETWORK MODEL\thanks{Work supported by
    U.S. Department of Energy}}

\author{J.-F. Ostiguy, K.-Y. Ng, FNAL, Batavia, IL 60510, USA}

\maketitle

\begin{abstract} 

During the course of the last decade, traveling wave accelerating structures for a future Linear Collider have been the object of
intense R\&D efforts. An important problem is the efficient computation of the long range wakefield 
with the ability to include small alignment and tuning errors. To that end, SLAC has developed an RF circuit model 
with a demonstrated ability to reproduce experimentally measured wakefields. 
The wakefield computation involves the repeated solution of a deterministic system of equations over a range of frequencies. 
By taking maximum advantage of the sparsity of the equations, we have achieved significant performance
improvements. These improvements make it practical to consider simulations involving an entire linac 
($\sim 10^3$ structures). One might also contemplate assessing, in real time, the impact of fabrication errors 
on the wakefield as an integral part of quality control.

\end{abstract}

\section{INTRODUCTION}
During the course of the last decade, SLAC has been conducting R\&D on 
new generations of accelerating structures for a future machine, the Next Linear Collider (NLC). 
The culmination of this work is the Damped Detuned Structure (DDS). Since it is difficult to dissipate deflecting 
mode power without also dissipating accelerating mode power, this structure achieves high efficiency (shunt impedance) 
by relying primarily on detuning to produce favorable phasing of the dipole modes to
mitigate the dipole sum wake. To prevent the partial re-coherence of the long range wake,
a small amount of damping is provided by extracting dipole mode energy through four manifolds which also serve as pumping slots.

A linear collider is a complex system and detailed numerical simulations 
are essential to understand the impact of different random and/or systematic 
structure fabrication errors on beam quality. Assuming a (loaded) gradient of 50 MV/m and a length of
2 m, each of the two arms of a 1 TeV in the center-of-mass NLC would be comprised of approximately 
$1000$ structures. To simulate the effect of fabrication errors on emittance growth, 
one needs to compute one wake per structure; consequently, there is considerable interest 
in performing these computations as efficiently as possible. A typical NLC structure comprises 
206 cells. Because of the large number of nodes, it impractical to resort to standard finite element or 
finite difference codes to compute the wake. To make computations manageable, the SLAC group has developed 
an RF circuit model. Despite its limitations, predictions have proven to be in remarkable 
agreement with experimental results. However, until now, the wake computations remained too slow to make the simulation 
of a full linac practical. In this paper, we describe algorithmic modifications that have led to a 
code achieving three orders of magnitude improvement over previously reported performance.

\section{CIRCUIT MODEL FOR DDS}
In an RF circuit model, Maxwell's equations are discretized using a low order expansion based on individual
closed cell modes. The result is a system of linear equations that can conveniently be represented by a circuit where voltages 
and currents are associated with modal expansion coefficient amplitudes.
A model suitable for the computation of the fields excited by the dipole excitation of a detuned structure
was developed by Bane and Gluckstern \cite{BANE}. The concept of manifold damping was later introduced by Kroll \cite{KROLL} 
and the circuit model was extended by the SLAC group to include this feature \cite{SPECTRAL}. 
The result is shown in Figure 1. The corresponding equations can be put in the form  
\begin{figure}[htb]
\centering
\includegraphics*[width=60mm,angle=270]{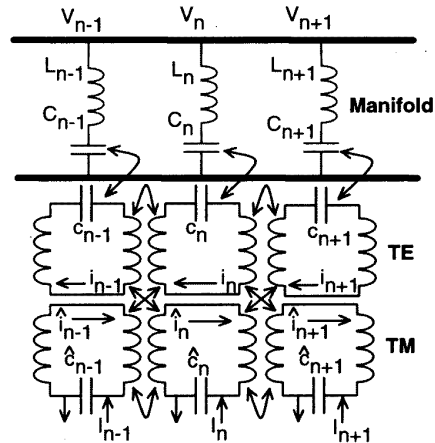}
\caption{Circuit model for Damped Detuned Structures. The thick horizontal lines represent a transmission line.}
\label{MOE05-f1}
\end{figure}
\begin{equation}
\left[ \begin{array}{ccc}
       H- \frac{1}{f^2} I    &    H_x                         &  0 \\
       H_x^T                   &   \hat{H} - \frac{1}{f^2} I  & -G \\
       0                       &    -G                          &  R 
       \end{array}
\right] 
\left[
       \begin{array}{c}
       a       \\
       \hat{a} \\
       A   
       \end{array}
\right]
= 
\frac{1}{f^2}
\left[
\begin{array}{c}
b \\
0  \\
0 
\end{array}
\right]
\label{e1}
\end{equation}
where $f$ is the frequency and $I$ is a unit diagonal. The submatrices $H$, $H_x$, $G$ and $R$ are $N \times N$ where $N$ is the number of cells (
$N=206$ for the SLAC structure). $H$ and $\hat{H}$ describe respectively the $\hbox{\rm TM}_{110}$-like and $\hbox{\rm TE}_{111}$-like cell 
mode coupling, $H_x$ represents the TE - TM cross coupling, $R$ describes the manifold mode propagation and $G$ describes the TE-to-manifold coupling. 
The vectors $a$, $\hat{a}$ are the normalized loop currents $( a = i/\sqrt{C_n})$ for the TM and TE chains and $V$ is
the normalized manifold voltage at each cell location. Finally, the right hand side $b$ represents the beam excitation. 
Since the boundary conditions at the cell interfaces impose that the TM and TE components must propagate in opposite directions, 
only the TM cell modes are excited by the beam. The dipole mode energy is coupled out electrically to the manifold via small slots; 
the TE component of the field is therefore capacitively coupled to the manifold. Note that the manifold is represented by a 
periodically loaded transmission line for which only nodal equations make sense, resulting in a mixed current-voltage formulation. 

%--------------------------------------------------------
\section{SPECTRAL FUNCTION}
%---------------------------------------------------------
Computing the wake of DDS structures involves solving (\ref{e1}) over the structure's dipole mode frequency bandwith.  
A longitudinal dipole impedance is first obtained by summing the cell voltages (in the frequency domain) with appropriate time delays. The  
transverse impedance is subsequently derived by invoking the Panofsky-Wentzel theorem. The circuit approach to wake computation introduces a small non-causal, non-physical component to the wake $w(t)$ which can be suppressed by considering $[w(t) - w(-t)] u_{-1}(t)$ instead. The sine transform of this function, proportional 
to the imaginary part of the impedance, is known as the spectral function $S(\omega)$. In the context of circuit-based wake computations, $S(\omega)$ 
is a more convenient quantity to compute than the dipole (beam) impedance. 

%--------------------------------------------------------
\section{SPARSE LINEAR EQUATIONS}
%---------------------------------------------------------

In the DDS circuit model, each cell couples only to its nearest neighbors. The resulting matrix is
sparse and complex symmetric (a consequence of electromagnetic reciprocity).
Computing the spectral function involves solving a sequence systems of linear equations. At each step in frequency,
the coefficient matrix changes slightly while its sparsity structure remains identical. In addition, a good starting 
approximation to the solution for any frequency step is provided by the solution from the previous step.  

\subsection{Iterative Methods}
%------------------------------

An algorithm suitable for symmetric complex systems is the so-called Quasi Minimal Residual (QMR) algorithm \cite{FREUND}.
This algorithm is a relative of the well-known conjugate gradient method which seeks to minimize the 
quadratic form $(Ax-b)^T \cdot (Ax-b)$. The QMR algorithm minizimizes a different quadratic form;
in both cases the key to rapid convergence is suitable ``preconditioning'' of the system $Ax=b$ with an approximate and 
easy to compute inverse.
Tests were performed with RDDS circuit matrices using standard incomplete factorization preconditioners; 
but the results were somewhat disappointing. It is believed that with a suitable preconditioner, the method can be competitive; 
however, efforts to identify one were abandoned after a direct technique proved to be more than satisfactory. 

\subsection{Direct Methods}
%----------------------------
Direct algorithms are essentially all relatives of the elementary Gaussian elimination algorithm, where unknowns are
eliminated systematically by linear combinations of rows. 

{\bf A crucial point is that the order in which the rows of the matrix are eliminated has a direct impact on 
computational efficiency since a different order implies different 
fill-in patterns \footnote{The elimination process creates non-zero entries at positions which correspond to zeros 
in the original coefficient matrix. The fill-in is the set of all entries which were originally zeros and took 
on non-zeros value at any step of the elimination process.}.}
In principle, there exists an elimination order that minimizes fill-in, which is {\bf not} the same as the most
numerically stable ordering. In some cases, it is even possible to find an ordering that produces no fill-in at all. 
Although the determination of a truly optimal ordering is an NP-complete problem, 
it is possible using practical strategies to find orderings that result in significant computational savings. 
The most successful class of ordering strategies are so-called ``local'' strategies that seek to minimize fill-in at each 
step in the elimination process regardless of their impact at a later stage. 
%---------------------------------------
\subsubsection*{The Markowitz Algorithm}
%---------------------------------------- 
A good local ordering strategy is the Markowitz algorithm. Suppose Gaussian elimination has proceeded through the first $k$
stages. For each row $i$ in the active $(n-k)\times(n-k)$ submatrix, let $r_i^{(k)}$ denote the number of entries. Similarly, let 
$c_j^{(k)}$ be the number of entries in column $j$. The Markowitz criterion is to select as pivot the entry $a_{ij}^{(k)}$ from the
$(n-k)\times(n-k)$ submatrix that satisfies 
\begin{equation}
\min_{i,j} (r_i^{(k)} -1) (c_j^{(k)} -1)
\label{e69ll}
\end{equation}  
Using this entry as the pivot causes $(r_i^{(k)} -1) (c_j^{(k)} -1)$ entry modifications at step $k$. Not all these modifications  
will result in fill-in; therefore, the Markowitz criterion is actually an approximation to the choice of pivot which introduces the least fill-in. 

%-------------------------
\section{CODE DESCRIPTION}
%-------------------------
Our code is based on the spectral function method and uses Markowitz ordering to solve the
circuit equations in the frequency domain. Compared to the procedure outlined in \cite{SPECTRAL}, 
the following changes have been made: (1) The manifold voltage $A$ is not separately eliminated, in order to preserve sparsity. 
(2) Once the system (\ref{e1}) is solved, the loop currents are known and the cell voltages can be 
obtained by a simple matrix multiplication. {\bf There is therefore no need to form an inverse}\cite{FN698}. 

Two additional remarks are in order. The process of determining the Markowitz ordering can by itself be time-consuming; 
however, since {\bf the structure of the RDDS matrix remains the same at every step in frequency, the ordering needs to be determined only once}. 
The relative magnitudes of the equivalent circuit matrix entries do not change very significantly over the 
frequency band occupied by the dipole modes. {\bf This insures that the Markowitz ordering 
remains numerically stable for all frequency steps}.

Implementations of the Markowitz algorithm are widely available. We used SPARSE \cite{SPARSE}, a
C implementation that takes advantage of pointers to store the coefficient matrix as a two-dimensional linked list. 
To each non-zero entry corresponds a list node. Each node in turn points to structure which comprises the numerical 
value of the entry, its two-dimensional indices and a pointer to an updating function. 
A linked list makes sequential traversal of a row or a column of the matrix efficient; however,
random access is expensive. To update the matrix at each frequency step, we sequentially scan the entire list 
and call an update function by indirection using a pointer stored within each entry structure.

The RDDS circuit matrix is not only sparse, it is also symmetric. The SPARSE package does not exploit this structure 
because the standard elimination process destroys symmetry. We note that the Markowitz scheme can be extented 
in a way that preserves symmetry.

%------------------------------
\section{RESULTS}
%------------------------------

\begin{figure}[h]
\center{\includegraphics[width=60mm,angle=-90]{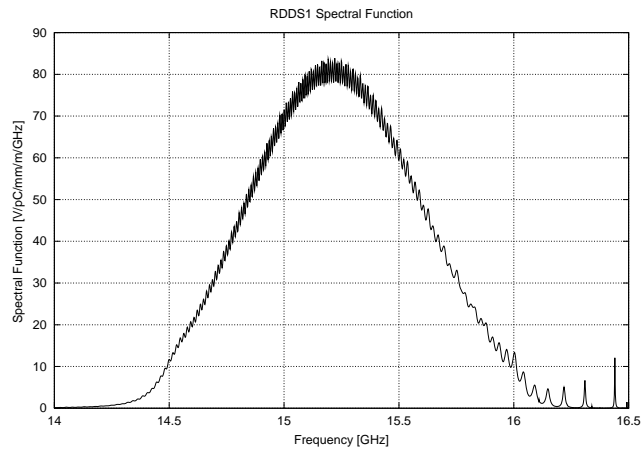}}
\caption{Computed spectral function for the RDDS1 structure.} 
\label{MOE05-f2}
\end{figure}

\begin{figure}[h]
\center{\includegraphics[width=60mm,angle=-90]{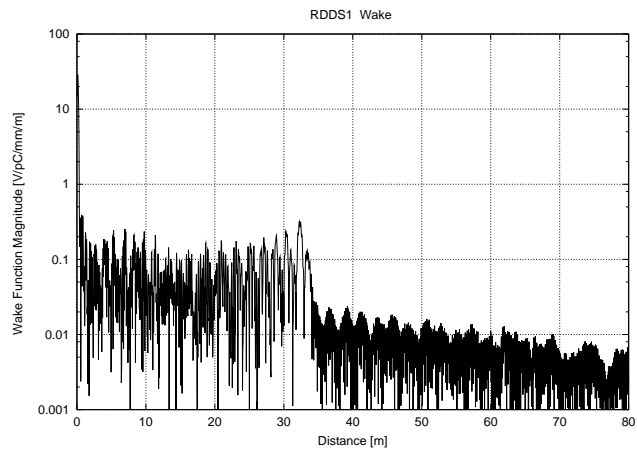}}
\caption{Computed wake for the RDDS1 structure.} 
\label{MOE05-f3}
\end{figure}

Our optimized wakefield code was used to compute the wake envelope of the RDDS structure, using parameters provided by SLAC. 
On a 550 MHz Pentium III (Linux, GNU gcc compiler) a complete calculation of the wake takes approximately 14 seconds. 
This represents a gain of roughly three orders of 
magnitude compared to the previously reported performance and allows the generation of wakes for an entire linac in
less than four hours. Output from the code is presented in Figures \ref{MOE05-f2} and \ref{MOE05-f3}. The results are identical to those 
obtained by the SLAC group. 
 

%--------------------
\section{ACKNOWLEDGMENTS}
%--------------------
The authors would like to express their appreciation to Norman Kroll, Roger Jones, 
Karl Bane, Roger Miller, Zhang-Hai Li and Tor Raubenheimer for in depth technical discussions 
about various aspects of the RDDS technology.  They also would like to extend special thanks 
to Norman Kroll and Roger Jones for generously sharing personal notes, providing 
parameters for the RDDS as well as sample wakefield computations.

\end{document}